\documentclass[twocolumn,showpacs,preprintnumbers,amsmath,amssymb,superscriptaddress]{revtex4}

\usepackage{graphicx}
\usepackage{dcolumn}
\usepackage{bm}

\begin{document}

\title{Ultrafast Optical Response of a High-Reflectivity GaAs/AlAs
Bragg Mirror
}

\author{Sara R. Hastings}
\affiliation{Department of Physics, University of California at
Santa Barbara, Santa Barbara, CA 93106 }

\author{Michiel J. A. de Dood}
\affiliation{Department of Physics, University of California at
Santa Barbara, Santa Barbara, CA 93106 }

\author{Hyochul Kim}
\affiliation{Department of Physics, University of California at
Santa Barbara, Santa Barbara, CA 93106 }

\author{William Marshall}
\affiliation{Department of Physics, University of Oxford, Oxford
OXI 3PU, United Kingdom } \affiliation{Department of Physics,
University of California at Santa Barbara, Santa Barbara, CA 93106
}

\author{Hagai S. Eisenberg}
\affiliation{Department of Physics, University of California at
Santa Barbara, Santa Barbara, CA 93106 }

\author{Dirk Bouwmeester}
\email[corresponding author: ]{bouwmeester@physics.ucsb.edu}
\affiliation{Department of Physics, University of California at
Santa Barbara, Santa Barbara, CA 93106 }


\begin{abstract}

The ultrafast response of a high-reflectivity GaAs/AlAs Bragg
mirror to optical pumping is investigated for all-optical
switching applications. Both Kerr and free carrier nonlinearities
are induced with 100~fs, 780~nm pulses with a fluence of
0.64~kJ/m$^2$ and 0.8~kJ/m$^2$. The absolute transmission of the
mirror at 931~nm increases by a factor of 27 from 0.0024\% to
0.065\% on a picosecond timescale. These results demonstrate the
potential for a high-reflectivity ultrafast switchable mirror for
quantum optics and optical communication applications. A design is
proposed for a structure to be pumped below the bandgaps of the
semiconductor mirror materials. Theoretical calculations on this
structure show switching ratios up to 2200 corresponding to
switching from 0.017\% to 37.4\% transmission.

\end{abstract}

\maketitle

High-finesse optical cavities are of interest in quantum optics
experiments, in particular for cavity quantum electrodynamics
\cite{Optical} and quantum state storage \cite{Marshall}. In many
of these experiments it would be beneficial to be able to switch
light in and out of a cavity on a fast timescale. Common cavity
switching techniques use intracavity elements which unavoidably
introduce additional cavity losses, limiting the finesse.  In
addition, switching elements such as acousto-optic modulators or
Pockels cells are limited to timescales longer than tens of
picoseconds.

Instead, we propose to switch the finesse of the cavity by
switching one of the cavity end mirrors.  The high-reflectivity
cavity mirrors are composed of alternating layers of two different
dielectric materials. Ideally the layer thicknesses in this Bragg
mirror are $\lambda/4n$, where $n$ is the refractive index of each
of the materials and $\lambda$ is the central wavelength of the
reflected light. If the index of refraction of at least one of the
materials can be switched rapidly, the reflectivity of the mirror
will change on the same time scale. The change in n alters the
ideal $\lambda/4n$ length ratio in the layers and the index
contrast between the two materials. This process can be used for
ultrafast all optical switching of a Bragg mirror
\cite{Eggleton,Taverner,Scalora,Hache}.

Similarly, switching in two and three dimensional photonic
crystals \cite{Leonard,Bristow,Mazurenko} and switching using
other mechanisms, such as spin-polarization relaxation
\cite{Nishikawa} and saturable absorption \cite{Bragg}, has been
studied.

This earlier work has focused primarily on switching by large
absolute percentages. However, to build a high-finesse switchable
cavity a mirror with high initial reflectivity a large switching
ratio is required. In this letter we present time resolved
pump-probe measurements of the change in transmission of a
GaAs/AlAs Bragg mirror under intense optical pumping.

A switchable mirror with high initial reflectivity requires
materials that have low absorption at the desired operation
wavelength, and a large index contrast is desirable in order to
keep the mirrors as thin as possible. At least one material must
posses a large nonlinear index of refraction to allow effective
all-optical switching. GaAs and AlAs meet these criteria and
mirrors with $\sim$30 layer pairs can be grown with reflectivities
$> $ 99.99\%. GaAs and AlAs have a Kerr nonlinearity and in
addition, the nonlinearity in index of refraction related to free
carriers in GaAs has previously been studied \cite{Huang}.

\begin{figure}
\includegraphics[width=8.6cm]{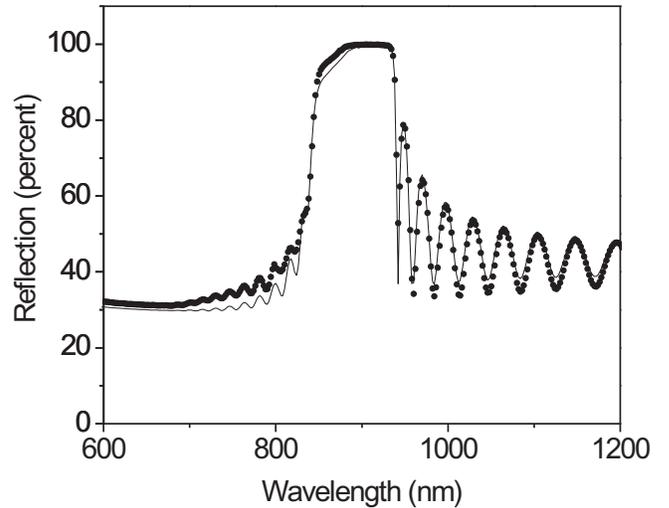} \caption{Measured reflectivity (circles) and calculated
reflectivity (solid line) of a 30 layer pair GaAs/AlAs Bragg
mirror at 12.5$^{\circ}$ angle of incidence. The mirror is
designed to have maximum reflectivity at 892nm for normal
incidence.} \label{wholemirror}
\end{figure}

The sample is a 30 pair GaAs/AlAs Bragg mirror on a GaAs substrate
with a $\sim$50~nm spacer layer of Al$_{0.4}$Ga$_{0.6}$As. The
thicknesses of the GaAs and AlAs layers are 61.8~nm and 75.0~nm
respectively, corresponding to $\lambda/4n$ for a wavelength of
892~nm. The measured reflectivity (circles) and calculated
reflectivity (solid line) as a function of wavelength at a
12.5$^\circ$ angle of incidence is shown in figure
\ref{wholemirror}.  The asymmetry in the reflectivity is caused by
absorption in the GaAs for photon energies larger than the bandgap
of the GaAs.

\begin{figure}
\includegraphics[width=8.6cm]{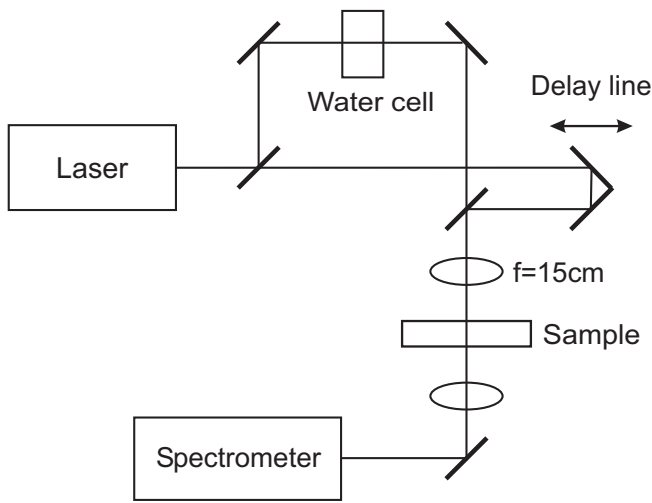} \caption{Setup
used to measure transmission through the mirror as a function of
temporal pump-probe overlap.  The probe is a broadband white light
created by continuum generation with part of the pump light. The
delay line in the pump path is scanned as transmission is measured
in a spectrometer.} \label{setup}
\end{figure}

The change in transmission through the sample as a function of the
delay between pump and probe pulses is studied using the setup
shown in figure \ref{setup}.  The light from a regeneratively
amplified titanium sapphire femtosecond mode-locked laser at 780
nm with $\sim$100 fs pulse width and 40 kHz repetition rate is
used as the pump. A portion of the light is split off and focused
into a cell of flowing water, generating ultrafast white light
probe pulses \cite{Fork}. The pump and probe are combined on a
dichroic mirror that reflects the 780nm pump beam and transmits
the white light probe for $\lambda$ $ > $ 820 nm such that they
propagate collinearly. The pump and probe are then focused to a 30
$\mu$m radius spot on the sample with a $f$=15 cm lens. The
collinearity of the pump and probe ensure good overlap on the
sample. The pump beam path has a delay line which is scanned and
at each position a spectrum of the transmitted light is measured
using a spectrometer with a cooled charged-coupled device (CCD)
camera. The pump light is absorbed in the sample, any residual
pump light is at a different wavelength from the probe and does
not interfere with the spectral measurement. A measurement of the
transmission demonstrates the ability to switch the light out of a
high-finesse cavity, as this requires a mirror that has an
increase in transmission under optical pumping.

\begin{figure}
\includegraphics[width=8.6cm]{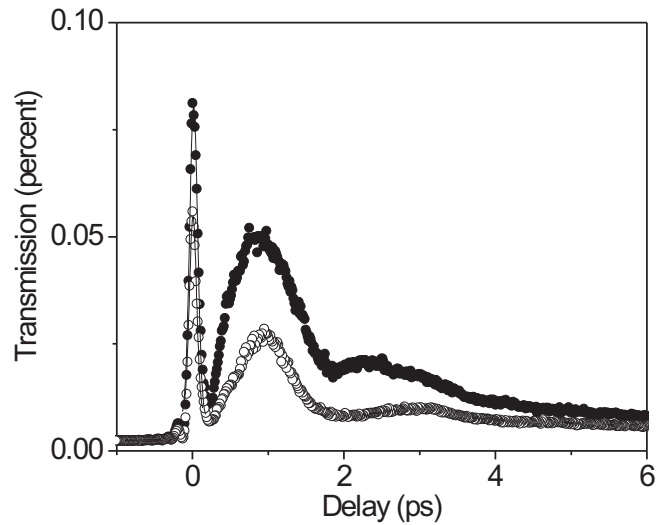}
\caption{Transmission at a wavelength of 931 nm as a function of
probe delay for a pump fluence 0.8 kJ/m$^2$ (solid circles) and
0.64 kJ/m$^2$ (open circles).} \label{singlewavelength}
\end{figure}

The transmission through the GaAs/AlAs mirror at 931nm for a pump
fluence of 0.8 kJ/m$^2$ (solid circles) and 0.64 kJ/m$^2$ (open
circles) as a function of pump probe delay is shown in figure
\ref{singlewavelength}. These fluences correspond to 80\% and 64\%
of the damage threshold for GaAs \cite{Huang}.  At negative delay
the transmission is constant. The initial fast response, peaking
at maximal pump probe overlap, is attributed to the Kerr
nonlinearity in GaAs and AlAs which changes the index of
refraction of both materials, leading to an increase in
transmission of the mirror. At 931nm this change is a 27 time
increase in transmission; from a transmission of 0.0024\% to
0.065\%. The first peak is fit to a Gaussian with a full width at
half maximum of $\sim$ 100 fs, consistent with the assumption that
the switching is due to an instantaneous (Kerr) nonlinearity. The
peak of the Gaussian corresponds to zero delay.

The second, lower but broader, peak is related to the presence of
free carriers that induce a change in the index of refraction and
increase the transmission of the mirror. Because the pump energy
is below the bandgap of AlAs, the free carriers are created
predominantly in the GaAs. A number of theoretical models for this
change in index of refraction have been introduced. For the
intense pump pulses used in our experiment, electrostatic
screening and many body effects from the large number of free
carriers are responsible for the index change \cite{Kim,Benedict}.

We also attribute the third, smaller peak after ~2.5 ps, to the
behavior of free carriers in the GaAs.  A detailed analysis would
require insight in the complicated dynamics of a high density of
free carriers in GaAs that interact with the lattice and is beyond
the scope of our experiments.  Thermal effects in GaAs are
typically observed on timescales $\sim$5 ps \cite{Huang}, and are
responsible for the small offset observed in Fig. 3 at 6 ps.

\begin{figure}
\includegraphics[width=8.6cm]{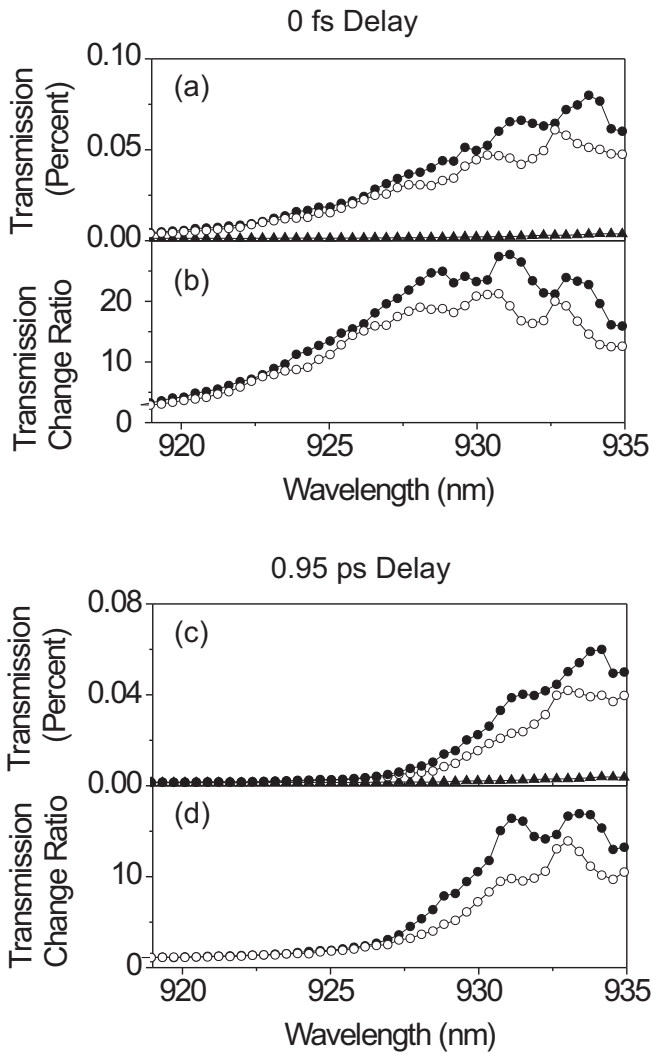}
\caption{Transmission spectrum for umpumped mirror (triangles) at
maximal pump probe overlap (a) and at the second peak in
transmission, a 0.95 fs delay (c) for pump fluence of 0.8 kJ/m$^2$
(solid circles) and 0.64 kJ/m$^2$ (open circles). The switching
occurs over the whole wavelength range measured. The ratio of the
pumped transmission to the umpumped transmission at zero delay (b)
and 0.95 fs delay (d). At zero delay the largest change occurs for
pump fluence of 0.8 kJ/m$^2$ at 931nm, a ratio of 27.  At a 945 fs
delay the largest change occurs for pump fluence of 0.8 kJ/m$^2$
at 933nm, a ratio of 17.} \label{spectrum}
\end{figure}

Figure \ref{spectrum}a shows the transmission as a function of
wavelength for the unpumped mirror (solid triangles), the mirror
at zero delay for a pump fluence of 0.8 kJ/m$^2$ (solid circles)
and 0.64 kJ/m$^2$ (open circles). The ratio of the transmission in
the pumped versus unpumped state is shown in figure
\ref{spectrum}b and is largest for the longer wavelengths and at a
pump fluence of 0.8 kJ/m$^2$. The maximum change is a 27 time
increase in transmission at a wavelength of 931~nm. The ratio of
change is larger for the longer wavelengths, closer to the edge of
the stop band of the Bragg mirror. There are two mechanisms that
contribute to this effect. An overall change in the refractive
index of the layers shifts the center wavelength of the Bragg
mirror.  In addition, a reduced index contrast between the layers
narrows the width of the stop band of the Bragg mirror.

The transmission at 0.95 ps delay, corresponding to the second
peak in transmission, is shown in figure \ref{spectrum}c.  The
overall switching ratio (figure \ref{spectrum}d) is less than that
at zero delay with a maximum ratio of 17 and an absolute change
from 0.0032\% to 0.054\%.

The absorption of the pump in the sample is assumed to be linear
in the GaAs layers and negligible in the AlAs. With an absorption
coefficient of 1.5$\times$ $10^{4}$ cm$^{-1}$ at 780 nm the 1/e
point for absorption of the pump is after $\sim$11 layer pairs.
The different pump intensity in the different layers produces a
different change in index of refraction for each layer, only
switching the top layers of the mirror effectively. However, with
lower absorption, the pump would propagate further into the mirror
and the switching ratio would be much larger.

The observation of switching due to the Kerr nonlinearity in GaAs
and AlAs demonstrates the potential to achieve a large switching
ratio using a pump laser at an energy below the bandgap in GaAs.
At this energy there is no linear absorption in the layers.

A 2x2 transfer matrix model for the transmission of a 30 layer
pair GaAs/AlAs Bragg mirror with the substrate etched away is used
to calculate the switching ratio for a 0.8~kJ/m$^2$ pump at 1060
nm. A two-photon absorption coefficient, $\beta~=~23~cm/GW$,
\cite{Said} is used to calculate an absorption coefficient of the
pump of 1.8~$\times$~$10^{6}~m^{-1}$ for the incident intensity.
For the two-photon process, the point where the intensity drops to
1/e times the initial value is after after 22 layers,
significantly larger than the 11 layers for pumping above the band
gap. A nonlinear coefficient $n_{2}$ =
-6.6~$\times$~$10^{-13}$~cm$^{2}$/W \cite{Said} in the GaAs layers
is assumed, where we have taken into account the co-linear double
beam configuration of the pump and probe \cite{Chiao}. Using the
values above, a switching ratio of 2200 is calculated with
transmission changing from 0.017\% to 37.4\% at 915nm. No data is
available for the Kerr nonlinearity in AlAs, but the nonlinearity
in AlAs below the bandgap is expected to be at least an order of
magnitude smaller than that of GaAs according to the dependence of
$n_{2}$ on the bandgap at pump energies below the bandgap
\cite{Sheik-Bahae,Ait}.

Using the Kerr nonlinearity to switch the mirror gives accurate
control over switching times. Pump pulses in the range from tens
of femtoseconds to tens of picoseconds could be used to achieve
desired switching times.

In conclusion we have shown that the nonlinear index of refraction
in GaAs and AlAs can be used to create a high-reflectivity
GaAs/AlAs all optically switchable mirror.  Switching is
demonstrated with a maximum change of 27 times in transmission
from 0.0024\% to 0.065\% at 931 nm. With a larger switching ratio
such a mirror would make an excellent optical switch as one end
mirror of a high-Q cavity. A switching ratio of 2200 is predicted
for optical pumping at energies below the bandgap of the GaAs.

\begin{acknowledgments}

The authors thank W. Irvine and C. Simon for useful discussions.
This work was supported by NSF grant PHY-0334970 and DARPA grant
MDA972-01-1-0027. SH is supported by a NSF Graduate Fellowship.

\end{acknowledgments}

\bibliography{mirrorswitching}

\begin{thebibliography}{19}
\expandafter\ifx\csname natexlab\endcsname\relax\def\natexlab#1{#1}\fi
\expandafter\ifx\csname bibnamefont\endcsname\relax
  \def\bibnamefont#1{#1}\fi
\expandafter\ifx\csname bibfnamefont\endcsname\relax
  \def\bibfnamefont#1{#1}\fi
\expandafter\ifx\csname citenamefont\endcsname\relax
  \def\citenamefont#1{#1}\fi
\expandafter\ifx\csname url\endcsname\relax
  \def\url#1{\texttt{#1}}\fi
\expandafter\ifx\csname urlprefix\endcsname\relax\def\urlprefix{URL }\fi
\providecommand{\bibinfo}[2]{#2}
\providecommand{\eprint}[2][]{\url{#2}}

\bibitem[{\citenamefont{Rempe et~al.}(1991)\citenamefont{Rempe, Thompson,
  Brecha, Lee, and Kimble}}]{Optical}
\bibinfo{author}{\bibfnamefont{G.}~\bibnamefont{Rempe}},
  \bibinfo{author}{\bibfnamefont{R.~J.} \bibnamefont{Thompson}},
  \bibinfo{author}{\bibfnamefont{R.~J.} \bibnamefont{Brecha}},
  \bibinfo{author}{\bibfnamefont{W.~D.} \bibnamefont{Lee}}, \bibnamefont{and}
  \bibinfo{author}{\bibfnamefont{H.~J.} \bibnamefont{Kimble}},
  \bibinfo{journal}{Phys. Rev. Lett.} \textbf{\bibinfo{volume}{67}},
  \bibinfo{pages}{1727} (\bibinfo{year}{1991}).

\bibitem[{\citenamefont{Marshall et~al.}(2003)\citenamefont{Marshall, Simon,
  Penrose, and Bouwmeester}}]{Marshall}
\bibinfo{author}{\bibfnamefont{W.}~\bibnamefont{Marshall}},
  \bibinfo{author}{\bibfnamefont{C.}~\bibnamefont{Simon}},
  \bibinfo{author}{\bibfnamefont{R.}~\bibnamefont{Penrose}}, \bibnamefont{and}
  \bibinfo{author}{\bibfnamefont{D.}~\bibnamefont{Bouwmeester}},
  \bibinfo{journal}{Phys. Rev. Lett.} \textbf{\bibinfo{volume}{91}}
  (\bibinfo{year}{2003}).

\bibitem[{\citenamefont{Eggleton et~al.}(1997)\citenamefont{Eggleton, Slusher,
  Judkins, Stark, and Vengsarkar}}]{Eggleton}
\bibinfo{author}{\bibfnamefont{B.~J.} \bibnamefont{Eggleton}},
  \bibinfo{author}{\bibfnamefont{R.~E.} \bibnamefont{Slusher}},
  \bibinfo{author}{\bibfnamefont{J.~B.} \bibnamefont{Judkins}},
  \bibinfo{author}{\bibfnamefont{J.~B.} \bibnamefont{Stark}}, \bibnamefont{and}
  \bibinfo{author}{\bibfnamefont{A.~M.} \bibnamefont{Vengsarkar}},
  \bibinfo{journal}{Opt. Lett.} \textbf{\bibinfo{volume}{22}},
  \bibinfo{pages}{883} (\bibinfo{year}{1997}).

\bibitem[{\citenamefont{Taverner et~al.}(1998)\citenamefont{Taverner,
  Broderick, Richardson, Laming, and Ibsen}}]{Taverner}
\bibinfo{author}{\bibfnamefont{D.}~\bibnamefont{Taverner}},
  \bibinfo{author}{\bibfnamefont{N.~G.~R.} \bibnamefont{Broderick}},
  \bibinfo{author}{\bibfnamefont{D.~J.} \bibnamefont{Richardson}},
  \bibinfo{author}{\bibfnamefont{R.~I.} \bibnamefont{Laming}},
  \bibnamefont{and} \bibinfo{author}{\bibfnamefont{M.}~\bibnamefont{Ibsen}},
  \bibinfo{journal}{Opt. Lett.} \textbf{\bibinfo{volume}{23}},
  \bibinfo{pages}{328} (\bibinfo{year}{1998}).

\bibitem[{\citenamefont{Scalora et~al.}(1994)\citenamefont{Scalora, Dowling,
  Bowden, and Bloemer}}]{Scalora}
\bibinfo{author}{\bibfnamefont{M.}~\bibnamefont{Scalora}},
  \bibinfo{author}{\bibfnamefont{J.~P.} \bibnamefont{Dowling}},
  \bibinfo{author}{\bibfnamefont{C.~M.} \bibnamefont{Bowden}},
  \bibnamefont{and} \bibinfo{author}{\bibfnamefont{M.~J.}
  \bibnamefont{Bloemer}}, \bibinfo{journal}{Phys. Rev. Lett.}
  \textbf{\bibinfo{volume}{73}}, \bibinfo{pages}{1368} (\bibinfo{year}{1994}).

\bibitem[{\citenamefont{Hache and Bourgeois}(2000)}]{Hache}
\bibinfo{author}{\bibfnamefont{A.}~\bibnamefont{Hache}} \bibnamefont{and}
  \bibinfo{author}{\bibfnamefont{M.}~\bibnamefont{Bourgeois}},
  \bibinfo{journal}{Appl. Phys. Lett.} \textbf{\bibinfo{volume}{77}},
  \bibinfo{pages}{4089} (\bibinfo{year}{2000}).

\bibitem[{\citenamefont{Leonard et~al.}(2002)\citenamefont{Leonard, van Driel,
  Schilling, and Wehrspohn}}]{Leonard}
\bibinfo{author}{\bibfnamefont{S.~W.} \bibnamefont{Leonard}},
  \bibinfo{author}{\bibfnamefont{H.~M.} \bibnamefont{van Driel}},
  \bibinfo{author}{\bibfnamefont{J.}~\bibnamefont{Schilling}},
  \bibnamefont{and} \bibinfo{author}{\bibfnamefont{R.~B.}
  \bibnamefont{Wehrspohn}}, \bibinfo{journal}{Phys. Rev. B}
  \textbf{\bibinfo{volume}{66}}, \bibinfo{pages}{161102}
  (\bibinfo{year}{2002}).

\bibitem[{\citenamefont{Bristow et~al.}(2003)\citenamefont{Bristow, Wells, Fan,
  Fox, Skolnick, Whittaker, Tahraoui, Krauss, and Roberts}}]{Bristow}
\bibinfo{author}{\bibfnamefont{A.~D.} \bibnamefont{Bristow}},
  \bibinfo{author}{\bibfnamefont{J.~P.~R.} \bibnamefont{Wells}},
  \bibinfo{author}{\bibfnamefont{W.~H.} \bibnamefont{Fan}},
  \bibinfo{author}{\bibfnamefont{A.~M.} \bibnamefont{Fox}},
  \bibinfo{author}{\bibfnamefont{M.~S.} \bibnamefont{Skolnick}},
  \bibinfo{author}{\bibfnamefont{D.~M.} \bibnamefont{Whittaker}},
  \bibinfo{author}{\bibfnamefont{A.}~\bibnamefont{Tahraoui}},
  \bibinfo{author}{\bibfnamefont{T.~F.} \bibnamefont{Krauss}},
  \bibnamefont{and} \bibinfo{author}{\bibfnamefont{J.~S.}
  \bibnamefont{Roberts}}, \bibinfo{journal}{Appl. Phys. Lett.}
  \textbf{\bibinfo{volume}{83}}, \bibinfo{pages}{851} (\bibinfo{year}{2003}).

\bibitem[{\citenamefont{Mazurenko et~al.}(2003)\citenamefont{Mazurenko, Kerst,
  Dijkhuis, Akimov, Golubev, Kurdyukov, Pevtsov, and Sel'kin}}]{Mazurenko}
\bibinfo{author}{\bibfnamefont{D.~A.} \bibnamefont{Mazurenko}},
  \bibinfo{author}{\bibfnamefont{R.}~\bibnamefont{Kerst}},
  \bibinfo{author}{\bibfnamefont{J.~I.} \bibnamefont{Dijkhuis}},
  \bibinfo{author}{\bibfnamefont{A.~V.} \bibnamefont{Akimov}},
  \bibinfo{author}{\bibfnamefont{V.~G.} \bibnamefont{Golubev}},
  \bibinfo{author}{\bibfnamefont{D.~A.} \bibnamefont{Kurdyukov}},
  \bibinfo{author}{\bibfnamefont{A.~B.} \bibnamefont{Pevtsov}},
  \bibnamefont{and} \bibinfo{author}{\bibfnamefont{A.~V.}
  \bibnamefont{Sel'kin}}, \bibinfo{journal}{Phys. Rev. Lett.}
  \textbf{\bibinfo{volume}{91}}, \bibinfo{pages}{213903}
  (\bibinfo{year}{2003}).

\bibitem[{\citenamefont{Nishikawa et~al.}(1995)\citenamefont{Nishikawa,
  Tackeuchi, Nakamura, Muto, and Yokoyama}}]{Nishikawa}
\bibinfo{author}{\bibfnamefont{Y.}~\bibnamefont{Nishikawa}},
  \bibinfo{author}{\bibfnamefont{A.}~\bibnamefont{Tackeuchi}},
  \bibinfo{author}{\bibfnamefont{S.}~\bibnamefont{Nakamura}},
  \bibinfo{author}{\bibfnamefont{S.}~\bibnamefont{Muto}}, \bibnamefont{and}
  \bibinfo{author}{\bibfnamefont{N.}~\bibnamefont{Yokoyama}},
  \bibinfo{journal}{Appl. Phys. Lett.} \textbf{\bibinfo{volume}{66}},
  \bibinfo{pages}{839} (\bibinfo{year}{1995}).

\bibitem[{\citenamefont{Kim et~al.}(1989)\citenamefont{Kim, Garmire, Hummel,
  and Dapkus}}]{Bragg}
\bibinfo{author}{\bibfnamefont{B.~G.} \bibnamefont{Kim}},
  \bibinfo{author}{\bibfnamefont{E.}~\bibnamefont{Garmire}},
  \bibinfo{author}{\bibfnamefont{S.~G.} \bibnamefont{Hummel}},
  \bibnamefont{and} \bibinfo{author}{\bibfnamefont{P.~D.}
  \bibnamefont{Dapkus}}, \bibinfo{journal}{Appl. Phys. Lett.}
  \textbf{\bibinfo{volume}{54}}, \bibinfo{pages}{1095} (\bibinfo{year}{1989}).

\bibitem[{\citenamefont{Huang et~al.}(1998)\citenamefont{Huang, Callan, Glezer,
  and Mazur}}]{Huang}
\bibinfo{author}{\bibfnamefont{L.}~\bibnamefont{Huang}},
  \bibinfo{author}{\bibfnamefont{J.~P.} \bibnamefont{Callan}},
  \bibinfo{author}{\bibfnamefont{E.~N.} \bibnamefont{Glezer}},
  \bibnamefont{and} \bibinfo{author}{\bibfnamefont{E.}~\bibnamefont{Mazur}},
  \bibinfo{journal}{Phys. Rev. Lett.} \textbf{\bibinfo{volume}{80}},
  \bibinfo{pages}{185} (\bibinfo{year}{1998}).

\bibitem[{\citenamefont{Fork et~al.}(1983)\citenamefont{Fork, Shank, Hirlimann,
  Yen, and Tomlinson}}]{Fork}
\bibinfo{author}{\bibfnamefont{R.~L.} \bibnamefont{Fork}},
  \bibinfo{author}{\bibfnamefont{C.~V.} \bibnamefont{Shank}},
  \bibinfo{author}{\bibfnamefont{C.}~\bibnamefont{Hirlimann}},
  \bibinfo{author}{\bibfnamefont{R.}~\bibnamefont{Yen}}, \bibnamefont{and}
  \bibinfo{author}{\bibfnamefont{W.~J.} \bibnamefont{Tomlinson}},
  \bibinfo{journal}{Opt. Lett.} \textbf{\bibinfo{volume}{8}},
  \bibinfo{pages}{1} (\bibinfo{year}{1983}).

\bibitem[{\citenamefont{Kim et~al.}(1994)\citenamefont{Kim, Ehrenreich, and
  Runge}}]{Kim}
\bibinfo{author}{\bibfnamefont{D.~H.} \bibnamefont{Kim}},
  \bibinfo{author}{\bibfnamefont{H.}~\bibnamefont{Ehrenreich}},
  \bibnamefont{and} \bibinfo{author}{\bibfnamefont{E.}~\bibnamefont{Runge}},
  \bibinfo{journal}{Solid State Communications} \textbf{\bibinfo{volume}{89}},
  \bibinfo{pages}{119} (\bibinfo{year}{1994}).

\bibitem[{\citenamefont{Benedict}(2001)}]{Benedict}
\bibinfo{author}{\bibfnamefont{L.~X.} \bibnamefont{Benedict}},
  \bibinfo{journal}{Phys. Rev. B} \textbf{\bibinfo{volume}{63}}
  (\bibinfo{year}{2001}).

\bibitem[{\citenamefont{Said et~al.}(1992)\citenamefont{Said, Sheik-Bahae,
  Hagan, Wei, Wang, Young, and Van~Stryland}}]{Said}
\bibinfo{author}{\bibfnamefont{A.~A.} \bibnamefont{Said}},
  \bibinfo{author}{\bibfnamefont{M.}~\bibnamefont{Sheik-Bahae}},
  \bibinfo{author}{\bibfnamefont{D.~J.} \bibnamefont{Hagan}},
  \bibinfo{author}{\bibfnamefont{T.~H.} \bibnamefont{Wei}},
  \bibinfo{author}{\bibfnamefont{J.}~\bibnamefont{Wang}},
  \bibinfo{author}{\bibfnamefont{J.}~\bibnamefont{Young}}, \bibnamefont{and}
  \bibinfo{author}{\bibfnamefont{E.~W.} \bibnamefont{Van~Stryland}},
  \bibinfo{journal}{J. Opt. Soc. Am. B} \textbf{\bibinfo{volume}{9}},
  \bibinfo{pages}{405} (\bibinfo{year}{1992}).

\bibitem[{\citenamefont{Chiao et~al.}(1966)\citenamefont{Chiao, Kelley, and
  Garmire}}]{Chiao}
\bibinfo{author}{\bibfnamefont{R.~Y.} \bibnamefont{Chiao}},
  \bibinfo{author}{\bibfnamefont{P.~L.} \bibnamefont{Kelley}},
  \bibnamefont{and} \bibinfo{author}{\bibfnamefont{E.}~\bibnamefont{Garmire}},
  \bibinfo{journal}{Phys. Rev. Lett.} \textbf{\bibinfo{volume}{17}},
  \bibinfo{pages}{1158} (\bibinfo{year}{1966}).

\bibitem[{\citenamefont{Sheik-Bahae et~al.}(1990)\citenamefont{Sheik-Bahae,
  Hagan, and Van~Stryland}}]{Sheik-Bahae}
\bibinfo{author}{\bibfnamefont{M.}~\bibnamefont{Sheik-Bahae}},
  \bibinfo{author}{\bibfnamefont{D.~J.} \bibnamefont{Hagan}}, \bibnamefont{and}
  \bibinfo{author}{\bibfnamefont{E.~W.} \bibnamefont{Van~Stryland}},
  \bibinfo{journal}{Phys. Rev. Lett.} \textbf{\bibinfo{volume}{65}},
  \bibinfo{pages}{96} (\bibinfo{year}{1990}).

\bibitem[{\citenamefont{Aitchison et~al.}(1997)\citenamefont{Aitchison,
  Hutchings, Kang, Stegeman, and Villeneuve}}]{Ait}
\bibinfo{author}{\bibfnamefont{J.~S.} \bibnamefont{Aitchison}},
  \bibinfo{author}{\bibfnamefont{D.~C.} \bibnamefont{Hutchings}},
  \bibinfo{author}{\bibfnamefont{J.~U.} \bibnamefont{Kang}},
  \bibinfo{author}{\bibfnamefont{G.~I.} \bibnamefont{Stegeman}},
  \bibnamefont{and}
  \bibinfo{author}{\bibfnamefont{A.}~\bibnamefont{Villeneuve}},
  \bibinfo{journal}{IEEE J. Quant. Elec.} \textbf{\bibinfo{volume}{33}},
  \bibinfo{pages}{341} (\bibinfo{year}{1997}).

\end{thebibliography}

\end{document}